\documentclass[12pt]{iopart}
\usepackage{graphicx}          
\usepackage{cite}              

\newcommand{\toot}{%
{\raisebox{0.1ex}
{\tiny$\left(\mbox{2\hspace{0.7ex}1}\atop\mbox{0\hspace{0.7ex}3}\right)$}}}
\newcommand{\toos}{%
{\raisebox{0.1ex}
{\tiny$\left(\mbox{2\hspace{0.7ex}1}\atop\mbox{0\hspace{0.7ex}6}\right)$}}}
\newcommand{\ttmoo}{%
{\raisebox{0.1ex}
{\tiny$\left(\mbox{~3\hspace{0.7ex}2}\atop\mbox{-1\hspace{0.7ex}1}\right)$}}}
\newcommand{\tootcaption}{%
{\raisebox{0.1ex}
{\small$\left(\mbox{2\hspace{0.7ex}1}\atop\mbox{0\hspace{0.7ex}3}\right)$}}}
\newcommand{\tooscaption}{%
\raisebox{0.1ex}
{\small$\left(\mbox{2\hspace{0.7ex}1}\atop\mbox{0\hspace{0.7ex}6}\right)$}}
\newcommand{\mc}[1]{\multicolumn{2}{c}{#1}}

\begin{document}


\title[Pb/Ge(001): Low-temperature two-dimensional phase transition]
  {Ge(001)-{\tootcaption}-Pb $\leftrightarrow$ {\tooscaption}-Pb:
   Low-temperature two-dimensional phase transition}

\author{O Bunk\dag\ddag\footnote[3]{To whom correspondence should be addressed:
  Oliver.Bunk@desy.de},
  M M Nielsen\ddag, J H Zeysing\dag, G Falkenberg\dag, F Berg-Rasmussen\ddag,
  M Nielsen\ddag, C Kumpf\ddag$\|$, Y Su\dag,\\
  R Feidenhans'l\ddag{} and R L Johnson\dag}
\address{{\dag} II. Institut f\"ur Experimentalphysik,
  Universit\"at Hamburg, Luruper Chaussee 149,
  D-22761 Hamburg, Germany}
\address{{\ddag} Condensed Matter Physics and Chemistry Department,
  Ris\o\ National Laboratory, DK-4000 Roskilde, Denmark}
\address{{$\|$} present address:  
Experimentelle Physik II, Universit\"at W\"urzburg, Am Hubland, 
D-97074 W\"urzburg, Germany}

\begin{abstract}
The Ge(001)-{\toot}-Pb surface reconstruction 
with a lead coverage of 5/3~monolayer is on the borderline
between the low-coverage covalently-bonded and high-coverage
metallic lead overlayers.
This gives rise to an unusual low-temperature phase transition 
with concomitant changes in the bonding configuration.
Both the room-temperature and low-temperature phases of this 
system were investigated by surface x-ray diffraction using 
synchrotron radiation.
The room-temperature Ge(001)-{\toot} phase is best described by
a model with dynamically flipping germanium dimers underneath a distorted
Pb(111) overlayer with predominantly metallic properties.
In the low-temperature Ge(001)-{\toos} phase the dimers are static
and the interaction between adsorbate and substrate and within 
the adsorbate is stronger than at room temperature. 
These results suggest that the phase transition is of order-disorder 
type. 
\end{abstract}


\vspace{28pt plus 10pt minus 18pt}
  \noindent{\small\rm Published in the {\it New Journal of Physics}
            \verb!http://www.njp.org!\par}

\maketitle

\section{Introduction}

Surfaces and interfaces determine many chemical and
physical properties of materials and play a key role 
in nanoscale electronic devices.
A well-known example of the influence of an interface
on electronic properties is the Schottky barrier
at metal-semiconductor interfaces. In principle the height
of the barrier is determined by the difference between the
work function of the metal and the electron affinity of the
semiconductor, but experimental investigations showed 
that the atomic structure directly at the interface may
influence the barrier height considerably 
\cite{moench_schottky_jvacscib:1999}.
Model systems chosen to study this phenomenon were, e.g., 
lead adsorbed on different silicon and germanium
surfaces, since for these systems abrupt interfaces with planar layers 
free of alloying were expected. 
Extensive scanning tunneling microscopy (STM) investigations
of the different phases of lead on Ge(001) surfaces for different lead
coverages and preparation temperatures partially confirmed
and partially disproved this simple idea 
\cite{yang_ge001pb_prb:1995,falkenberg_pbge001_surfsci:1997}. 
For lead coverages above $\sim$1.5~monolayer (ML) 
quite flat and uniform lead overlayers forming a
commensurate {\toot} and at higher coverage an
incommensurate c(8$\times$4)i reconstruction are observed. 
That the latter overlayer exhibits metallic properties is evident from
the low STM tip bias of 20~mV used to image this
surface\cite{falkenberg_pbge001_surfsci:1997}. 
Below 1.5~ML lead coverage and at preparation temperatures
below 300$^\circ$C commensurate reconstructions with
(2$\times$2) and c(8$\times$4) unit-cells are formed. Both
reconstructions are described by lead dimers covalently
bonded to the germanium substrate. For the
Ge(001)-(2$\times$2)-Pb reconstruction with a nominal lead coverage of 
0.5~ML the germanium substrate is still dimerized whereas in the 
c(8$\times$4) phase with a nominal lead coverage of 0.75~ML 
half the dimer bonds are broken \cite{falkenberg_pbge001_surfsci:1997} 
-- as confirmed recently by first-principle total energy calculations 
\cite{takeuchi_ge0012x2pb_surfsci:1998,takeuchi_ge001c8x4pb_prb:1998}. 
The lead dimers in these two reconstructions are 
tilted to avoid the energetically unfavourable 
configuration of two approximately sp$^3$ hybridized 
lead atoms each with a half-filled ``dangling bond''. 
This buckling is static as opposed to the clean
Si(001)- and Ge(001)-(2$\times$1) surfaces where the dimers 
flip dynamically at room temperature
\cite{wolkow_si0012x1_prl:1992}. 
The lead-induced reconstructions of the Ge(001)
surface described so far more or less exhibit the expected
flat overlayer necessary for an ideal Schottky barrier model
system -- although only the phases with a sufficiently high lead coverage 
fulfill the condition of being metallic. 
At low lead coverages below 1~ML and preparation
temperatures above $\sim$300$^\circ$C the surface exhibits a
completely different morphology. This phase is not well
ordered and has been described as the ``rough phase''
\cite{falkenberg_pbge001_surfsci:1997} with height
differences of more than five atomic layers. The formation of this 
phase with an unusually high roughness for lead overlayers may
be explained by the significant stress induced by the
incorporation of lead atoms with a covalent radius of
1.47~\AA{} on germanium lattice sites (covalent radius
1.22~\AA{}), i.e., an unexpected alloying may occur at low
lead coverages and high preparation temperatures. 

The lead-induced surface reconstructions of the
Ge(001) surface described so far possess a wide diversity 
of properties. But the Ge(001)-{\toot}-Pb reconstruction, 
formed at an intermediate lead coverage of around 5/3~ML
seems to be even more interesting. From the point of view of
coverage it resides between the low-coverage 
covalently-bonded and the high-coverage metallic
phases and may share properties common to both. 

Based on thorough STM investigations and some diffraction results 
a structural model was proposed 
for the Ge(001)-{\toot}-Pb reconstruction with 5/3~ML lead 
in a distorted Pb(111) layer on top of a completely dimerized 
Ge(001) substrate 
\cite{falkenberg_pbge001_surfsci:1997,falkenberg_thesis:1998}. 
The complete dimerization of the germanium substrate is surprising 
since at lower lead coverage in the Ge(001)-c(8$\times$4)-Pb 
reconstruction the dimerization of the 
(even lower coverage) Ge(001)-(2$\times$2)-Pb surface and the 
clean Ge(001)-(2$\times$1) is already partially removed. 
To verify the proposed structural model for the room-temperature phase 
and to elucidate the properties of the novel 
Ge(001)-{\toos}-Pb low-temperature phase 
\cite{jahns_phd:1994} we investigated 
both systems in a surface x-ray diffraction (SXRD) experiment using 
synchrotron radiation.

\section{Experimental}

\subsection{Preparation and measurements}

The sample was prepared in an ultra-high vacuum (UHV) system equipped with 
reflection high-energy electron diffraction (RHEED) and 
low-energy electron diffraction (LEED) facilities. 
A clean (2$\times$1) reconstructed Ge(001) sample was prepared by 
the standard procedure of cycles of 
sputtering at 350-450$^\circ$C for 30-60~min using 500~eV Ar$^+$ ions 
and annealing at 600-650$^\circ$C for 15~min. 
Lead was deposited while observing the RHEED-patterns: 
Starting from the (2$\times$1) reconstruction of the 
clean surface first the c(8$\times$4)-Pb and finally the 
Ge(001)-{\toot}-Pb phase was observed. 
The change from c(8$\times$4) to {\toot} periodicity is accompanied by the 
disappearance of the characteristic c(8$\times$4)-streaks and the emergence of 
half-order reflections in the $(1\overline{1})$ azimuth. 
The long-range order of the surface reconstruction was improved by annealing 
the sample at 150$^\circ$C. 

The sample was then transferred to a portable UHV chamber equipped with 
a helium closed-cycle two-stage sample cooling system and brought 
to the BW2 wiggler beamline at the 
Hamburg Synchrotron Radiation Laboratory (HASYLAB) 
for the x-ray diffraction measurements. 
The x-ray photon energy was set to 10~keV ($\lambda = 1.24$~\AA{}) 
and the glancing angle of incidence to 0.65$^\circ$ 
(i.e.\ above the critical angle to reduce 
the uncertainties in the measured intensities arising from mechanical 
displacements). 
A data-set consisting of 182 in-plane reflections, 
236 reflections along nine fractional-order rods and 81 reflections along 
three integer-order rods (so-called crystal truncation rods, CTRs) 
was recorded for the Ge(001)-{\toot}-Pb structure determination. 
After completing the room-temperature measurements the sample was cooled until
the temperature of the sample holder reached 20~K. 
For the low-temperature {\toos}-phase 313 in-plane reflections, 
371 reflections along 18 fractional-order rods and 106 reflections along 
four CTRs were recorded. 

\subsection{Data-analysis}

The integrated intensities were corrected for the Lorentz factor, 
polarization factor, active sample area and the rod interception 
appropriate for the z-axis geometry \cite{vlieg_correction_factors}. 
The in-plane data were averaged 
assuming that all four domains which are possible within the 
p2-symmetry are equally populated. 
In this way the room-temperature in-plane data set was reduced to 
96 and the low-temperature in-plane data set to 223 
symmetry-inequivalent reflections. 
The recording of extensive data sets with 413 symmetry-inequivalent 
data points for the room-temperature and 700 data points for the 
low-temperature phase was necessary to determine all the relevant parameters 
in these complex reconstructions with large unit cells. 

The conventional surface coordinate system with 
${\bf a} = 1/2\,[110]_{\mbox{cubic}}$, 
${\bf b} = 1/2\,[\overline{1}10]_{\mbox{cubic}}$ and
${\bf c} = [001]_{\mbox{cubic}}$ is used. The cubic coordinates are 
in units of the germanium lattice parameter of 5.66~\AA{} at 300~K and 
therefore $|{\bf a}| = |{\bf b}| = 4.00$~\AA{} and $|{\bf c}| = 5.66$~\AA{}. 

Only seven layers were included in the data-analysis 
to describe the substrate relaxations, i.e., 
less than the ten layers used for the analysis of 
the clean surface \cite{rossmann_ge0012x1_surfsci:1992} to keep the number 
of free parameters reasonably low. 
In addition the elastic energy calculated using the 
Keating algorithm \cite{keating_pr:1966,pedersen_keating_surfsci:1989}
was used as a constraint on the substrate parameters in the 
data analysis. 
The Keating term restricts the substrate parameters if the optimization 
runs into a shallow $\chi^2$ minimum. 
Nevertheless the data analysis was only possible using 
a computer program with the ability to handle efficiently the 152 parameters 
necessary to describe the atomic positions in the low-temperature phase 
together with further parameters describing thermal motion and 
disorder and the Keating constraints for 105 of the parameters. 

\begin{figure}
  \begin{center}
    {\large (a)}\includegraphics[width=5.0cm]{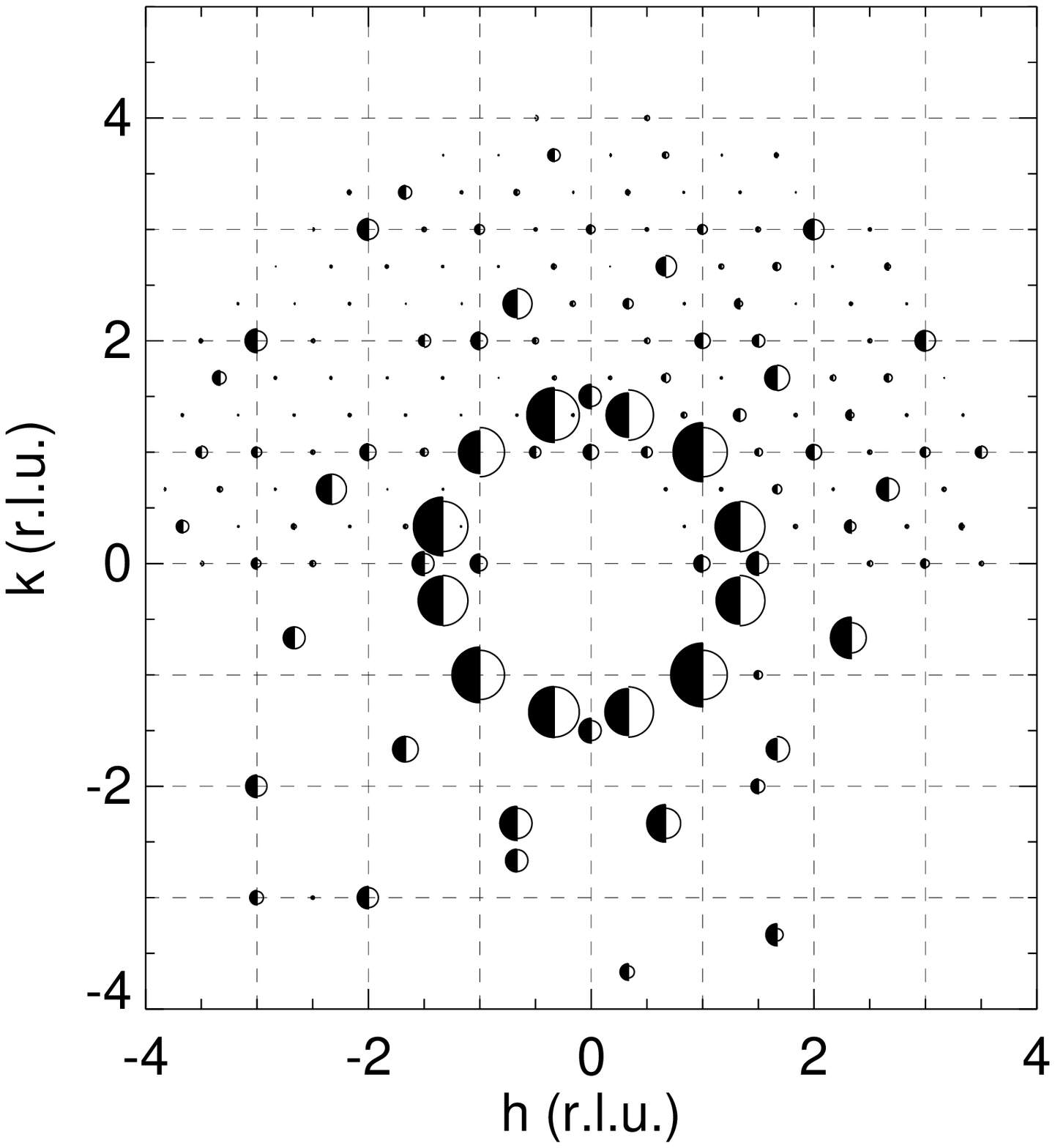}
    \vspace{3mm}

    {\large (b)} \includegraphics[width=8.0cm]{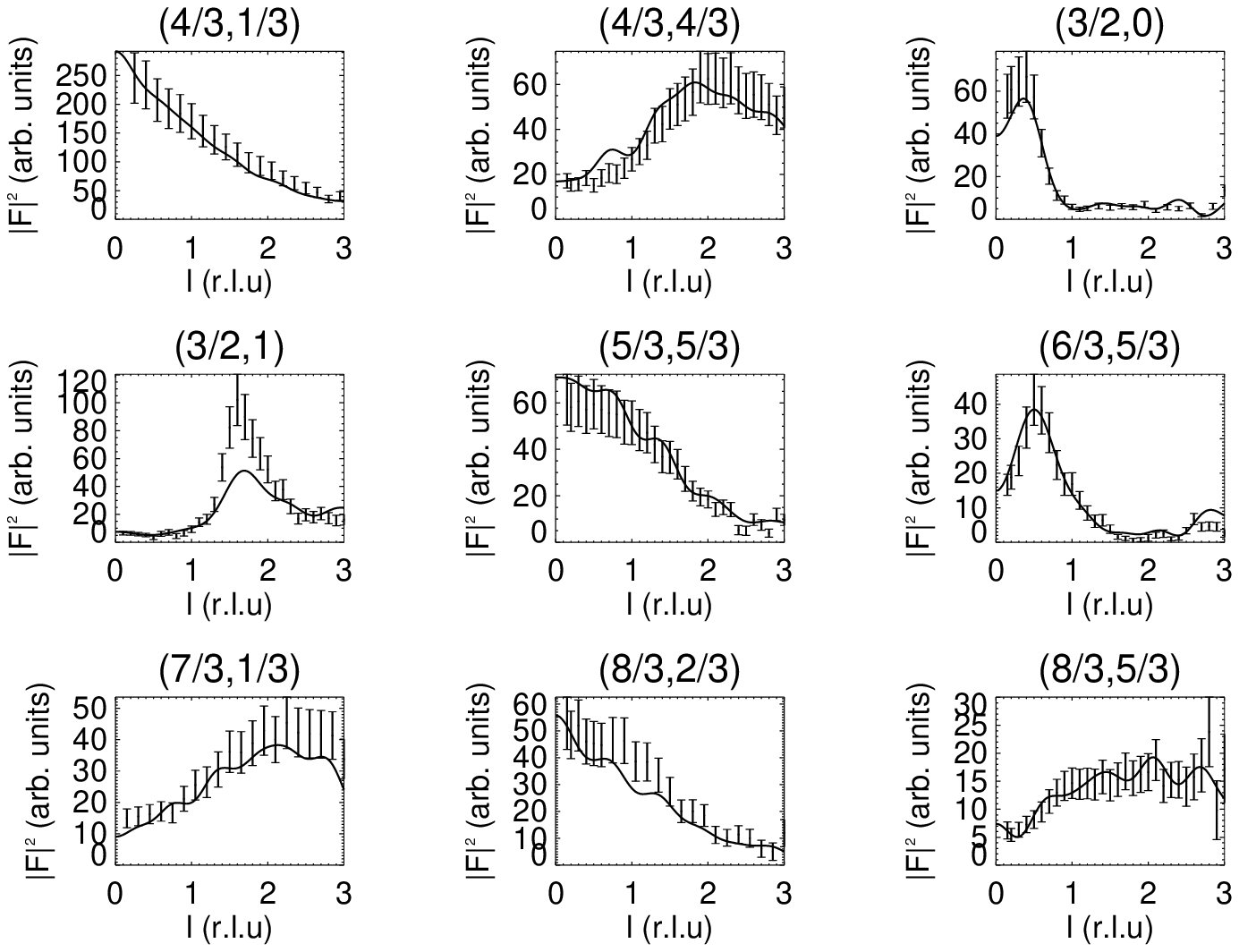}
    \vspace{3mm}

    {\large (c)} \includegraphics[width=8.0cm]{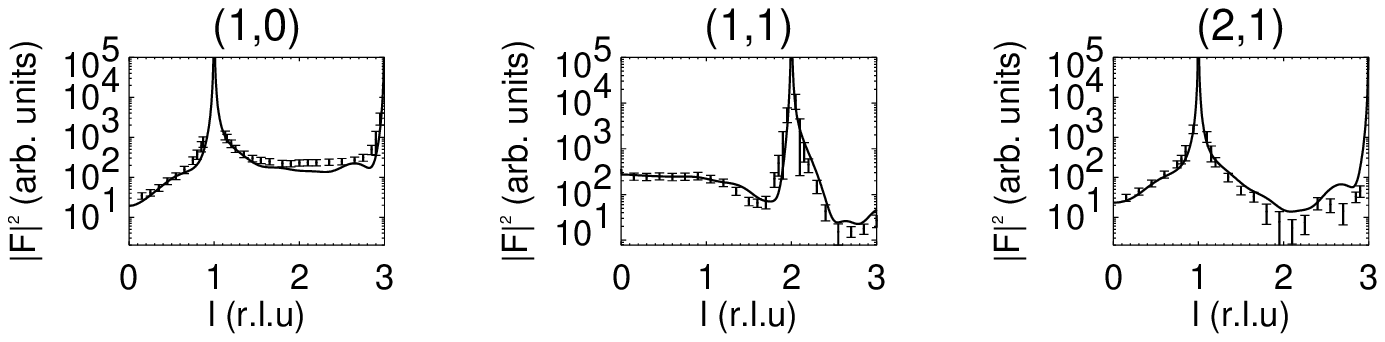}
  \end{center}
  \caption{SXRD data for the Ge(001)-{\toot}-Pb room-temperature phase. 
    (a) In-plane data measured at a small $l$-value of 0.13. 
    Unaveraged data are shown to display the characteristic pattern 
    of twelve reflections with high intensity which originates from 
    the lead overlayer with (approximately) sixfold symmetry on the 
    germanium substrate with twofold symmetry. 
    The areas of the filled (empty) semi-circles are proportional 
    to the measured (calculated) intensities. 
    (b) Fractional-order and (c) integer-order rod scans. 
    The solid line is calculated using 
    the model shown in \fref{fig:2103-model} with the parameters 
    given in \tref{tab:2103}.
    \label{fig:2103-data}}
\end{figure}

\begin{figure}
  \begin{center}
    {\large (a)}\includegraphics[width=6.0cm]{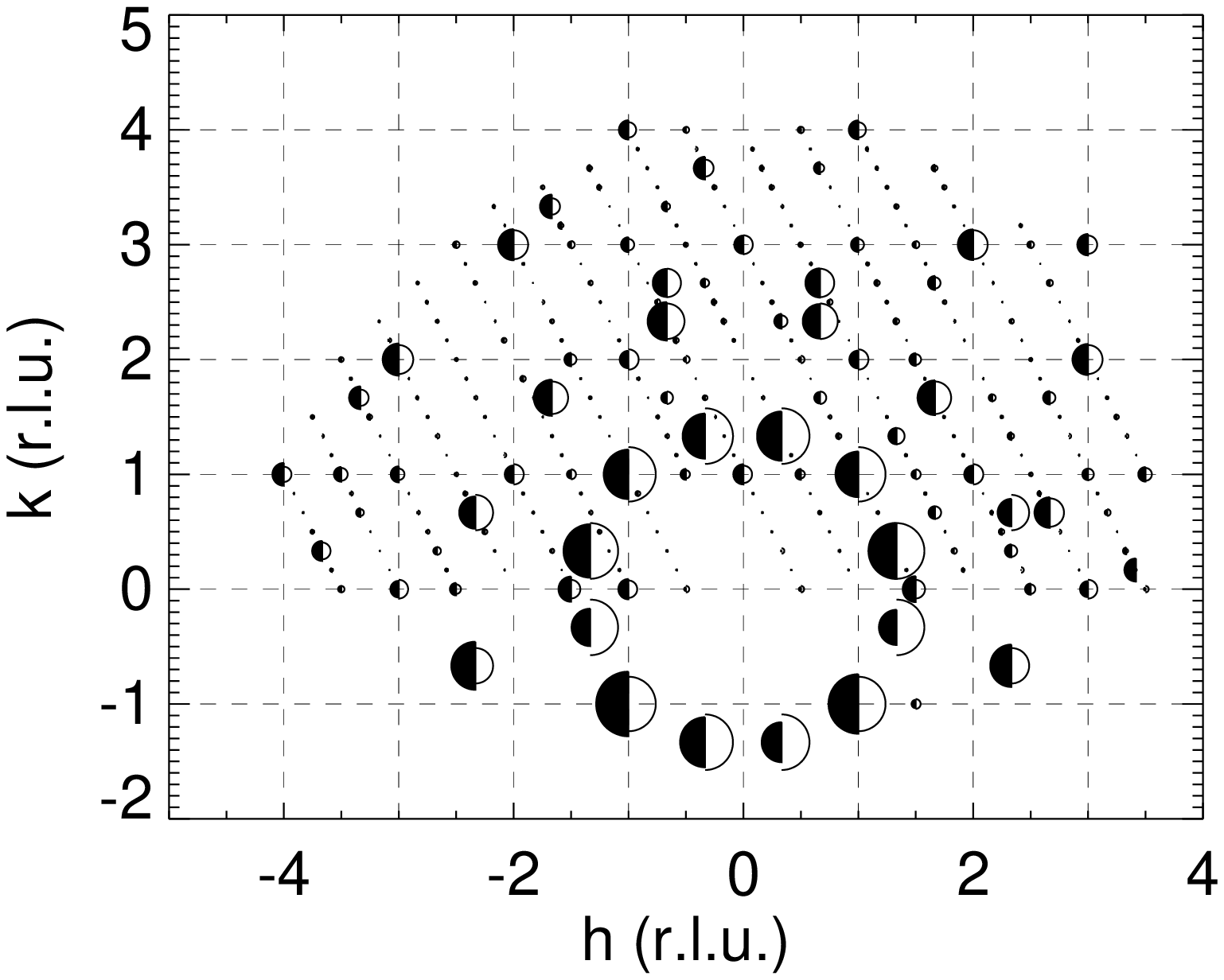}
    \hspace{1mm}
    {\large (b)} \includegraphics[width=5.5cm]{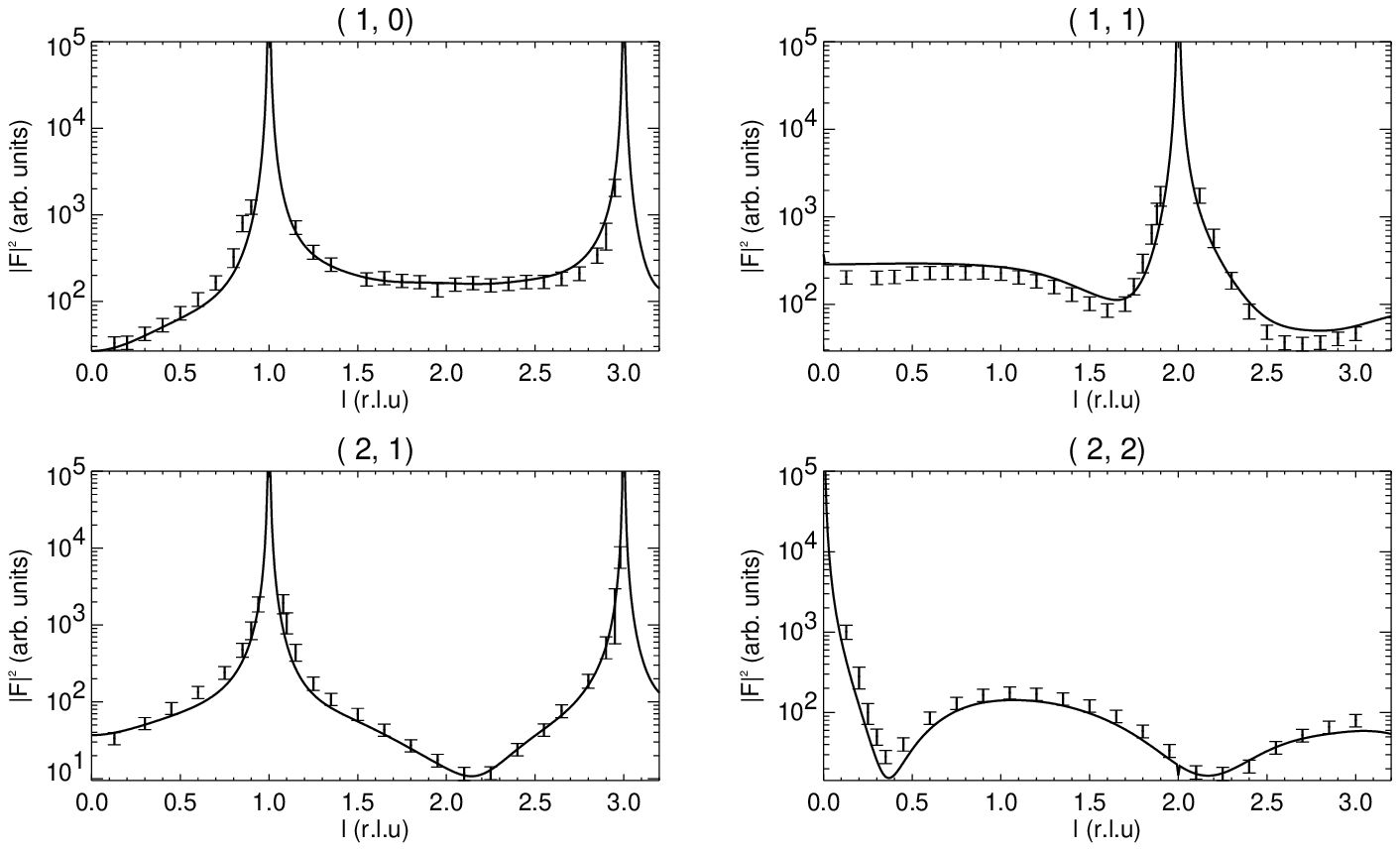}
    \vspace{3mm}

    {\large (c)} \includegraphics[width=7.0cm]{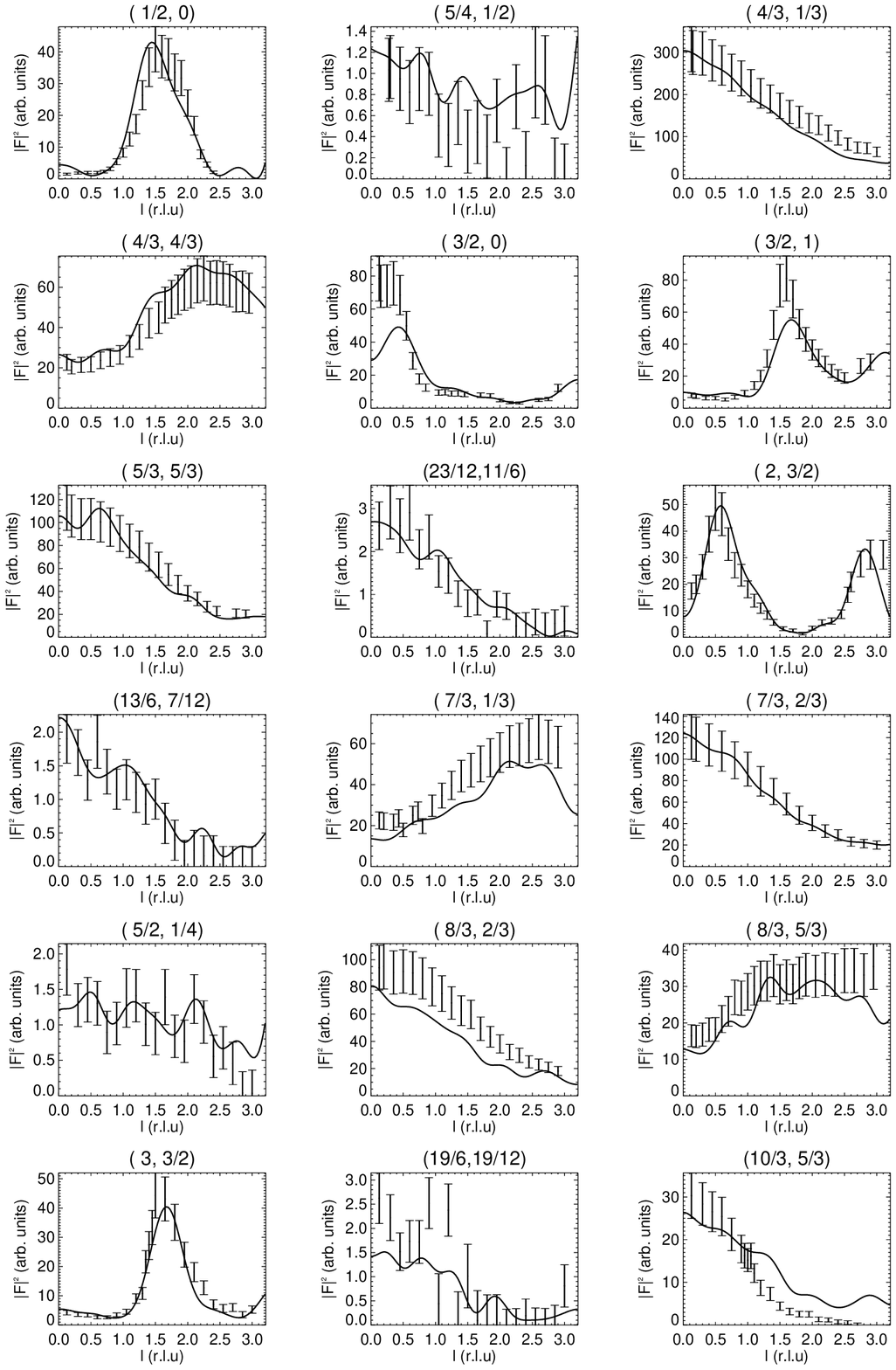}
  \end{center}
  \caption{SXRD data for the Ge(001)-{\toos}-Pb low-temperature phase. 
    (a) In-plane data measured at a small $l$-value of 0.13. 
    The areas of the filled (empty) semi-circles are proportional 
    to the measured (calculated) intensities. 
    (b) Integer-order and (c) fractional-order rod scans. 
    The solid line is calculated using 
    the model shown in \fref{fig:2106-model} with the parameters 
    given in \tref{tab:2106-1} and \tref{tab:2106-2}.
    \label{fig:2106-data}}
\end{figure}

The overall good agreement between the result of the structural
refinement and the measured data is apparent from
\fref{fig:2103-data} for the room-temperature results and 
\fref{fig:2106-data} for the low-temperature data. 
Using in total 82 parameters a reduced $\chi^2(|F|)$ value of 2.2 was
obtained for the room-temperature phase. 
For the low-temperature phase in total 155 parameters were
determined and the reduced $\chi^2(|F|)$ was 3.3. 
From a technical point of view the number of parameters is
exceptionally large for a surface structure determination. 

\section{Results and discussion}

In the following two subsections we focus on the most important
results and suggest a possible interpretation. 

\subsection{Room-temperature phase}

\begin{figure}
  \begin{center}
    \includegraphics[width=7cm]{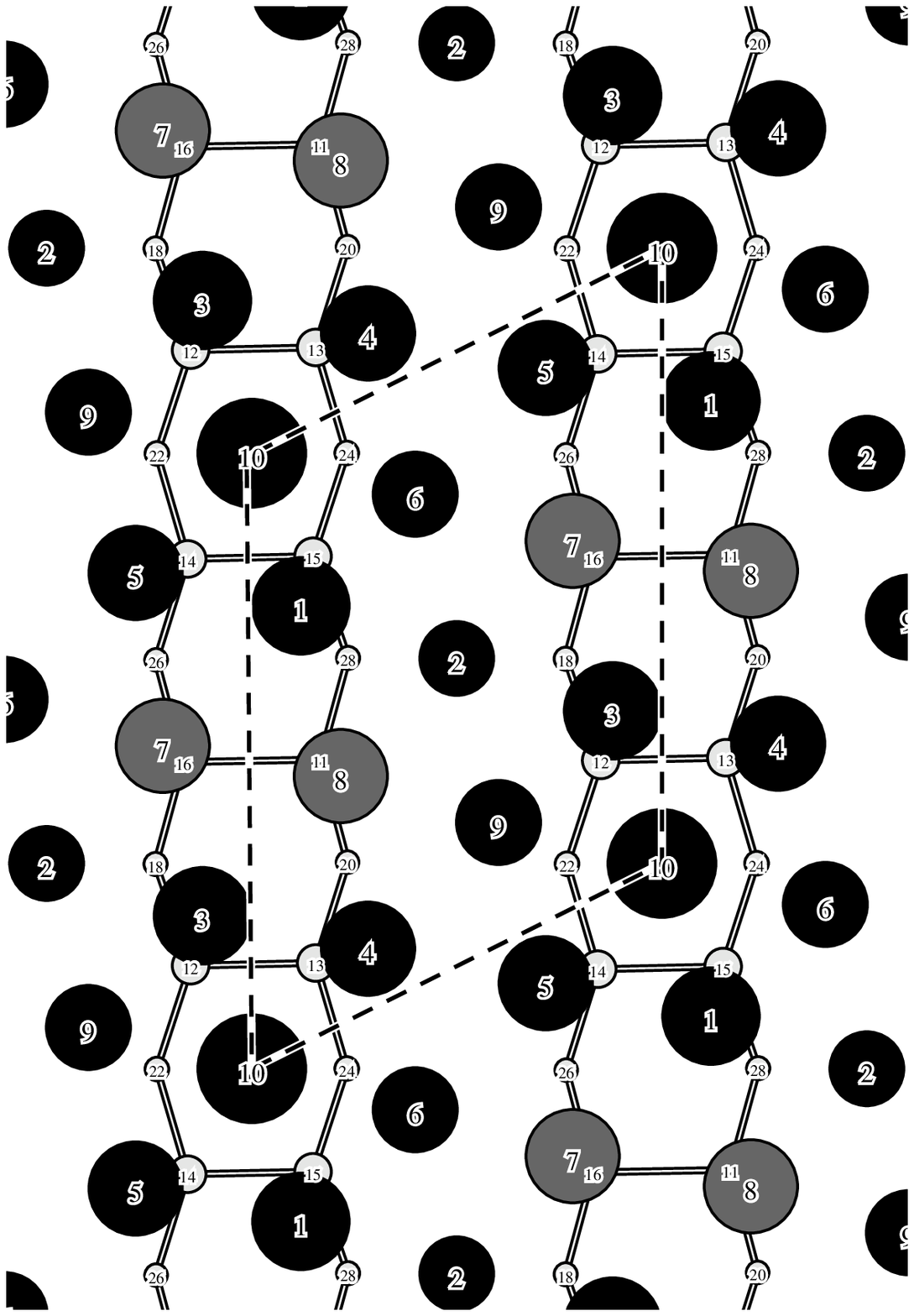}
  \end{center}
  \caption{Atomic structure of the Ge(001)-{\toot}-Pb room-temperature 
    phase. The {\toot} unit-cell is indicated by a dashed line. 
    Lead atoms are drawn black or dark grey. 
    The dark grey lead atoms are 2.7~\AA{} above the substrate. 
    Germanium atoms are the small circles. 
    The numbers shown correspond to \tref{tab:2103}. 
    Lead atoms 2 and 10 are on high-symmetry sites. For simplicity all 
    atoms have separate labels irrespective of the p2-symmetry restrictions.
    \label{fig:2103-model}}
\end{figure}

The analysis of the room-temperature data basically confirmed the
structural model with a distorted Pb(111) layer on top of the 
dimerized germanium substrate proposed previously by Falkenberg \etal 
\cite{falkenberg_pbge001_surfsci:1997} 
(see \fref{fig:2103-model} and \tref{tab:2103}). 
We could confirm that the germanium dimers of the clean surface 
broken when forming the Ge(001)-c(8$\times$4)-Pb reconstruction 
are reestablished at a higher lead coverage of 5/3~ML. 
In addition to this remarkable result some interesting details
are revealed. 

The atomic displacement parameter (ADP) used in the analysis of diffraction 
data describes both thermal motion and static disorder 
\cite{trueblood_adp_actacrysta:1996}. 
The ADP listed in \tref{tab:2103} for the lead atoms 1-9 
(see \fref{fig:2103-model}) is anisotropic -- larger in the
surface plane than perpendicular to it. 
This suggests the picture of a metallic overlayer with 
static disorder and a significant mobility of the individual
atoms around their equilibrium position. 
An extra ADP was necessary to describe the behavior 
of lead atom 10. 
This atom is on a high-symmetry site and has a larger distance 
to the germanium substrate than the other lead atoms. 
It is therefore more weakly bonded. 
This is reflected in its higher ADP. 

There is a striking anisotropy in the ADP of the first layer germanium 
atoms. The amplitude of $0.4\pm 0.1$~\AA{} in the $z$-direction 
(perpendicular to the surface plane) matches well to the values of 
0.50~\AA{} \cite{rossmann_ge0012x1_surfsci:1992} and 
0.42~\AA{} \cite{torrelles_ge0012x1_surfsci:1996} for the clean surface. 
This may indicate that, similar to the clean surface, 
the germanium dimers are dynamically flipping beneath the lead
overlayer. 
This analogy to the clean surface seems to be obvious, 
although static disorder and dynamic movement 
cannot be distinguished by SXRD. The bond length of $2.45 \pm 0.08$~\AA{} 
for two germanium dimers (12-13 and 14-15) is in excellent agreement 
with the value of $2.44 \pm 0.01$~\AA{} for the clean Ge(001)-(2$\times$1) 
reconstruction \cite{rossmann_ge0012x1_surfsci:1992}. 
The simple picture of an unperturbed germanium surface layer 
is supported by this fact. 
However, a bond length of $2.68 \pm 0.07$~\AA{} was determined for the 
germanium dimer 11-16. This is an increase of 9~\%{} compared to the 
substrate bond length of 2.45~\AA{}. Since the germanium atoms 11 and 16 
have the smallest observed Ge-Pb distance of $2.71 \pm 0.06$~\AA{} 
to their next neighbor atoms 7 and 8 
(drawn dark grey in \fref{fig:2103-model}) a stronger, partially covalent 
Pb-Ge interaction weakening the Ge-Ge dimer bonds may cause the observed 
increase in the dimer bond length. 

To summarize the results for the Ge(001)-{\toot}-Pb room-temperature phase, 
the following model is consistent with the observed structural parameters: 
A distorted, metallic Pb(111) overlayer is only weakly bonded 
within the layer and to the substrate 
and has an increased in-plane motion or disorder compared 
to the out-of-plane direction. 
A partially covalent interaction between one germanium dimer and the 
two next neighbor lead atoms per unit-cell stabilizes the
overlayer. 
This interaction tends to make the overlayer commensurate with
the substrate and to weaken the corresponding dimer bond under
the lead layer. 
The remaining two substrate dimers per unit-cell 
continue the dynamic flipping movement characteristic of the
clean surface. 
The delicate balance between covalent and metallic behavior of the surface 
atoms indicates that this system is likely to respond to small changes 
of the total energy, e.g., by simply changing the temperature. 
The structural parameters derived from the SXRD measurements 
suggest that this system is predisposed to structural changes as
a function of the temperature. 

\subsection{Low-temperature phase}

\begin{figure}
  \begin{center}
    {\large (a)}\includegraphics[width=9.5cm]{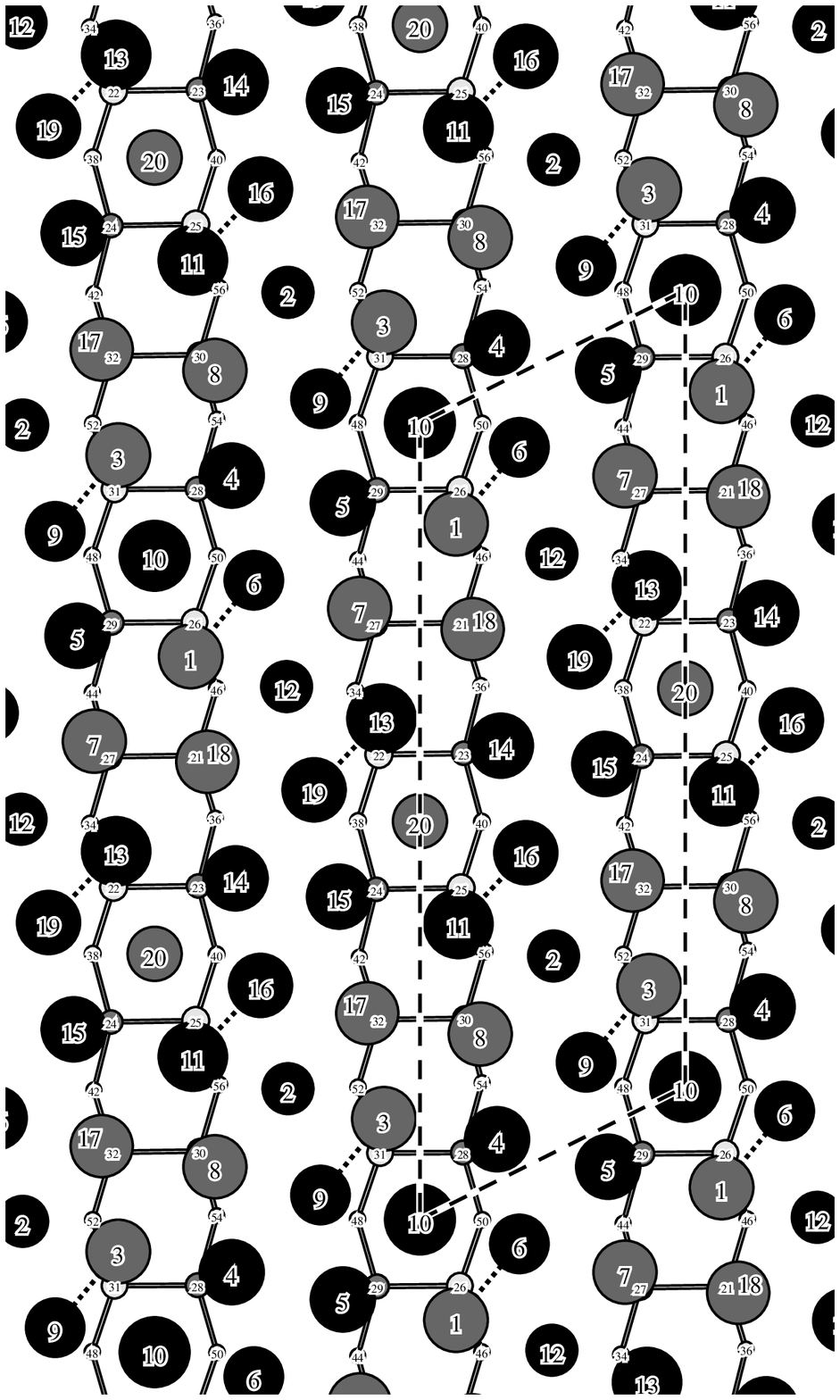}
    {\large (b)}\includegraphics[width=8.0cm]{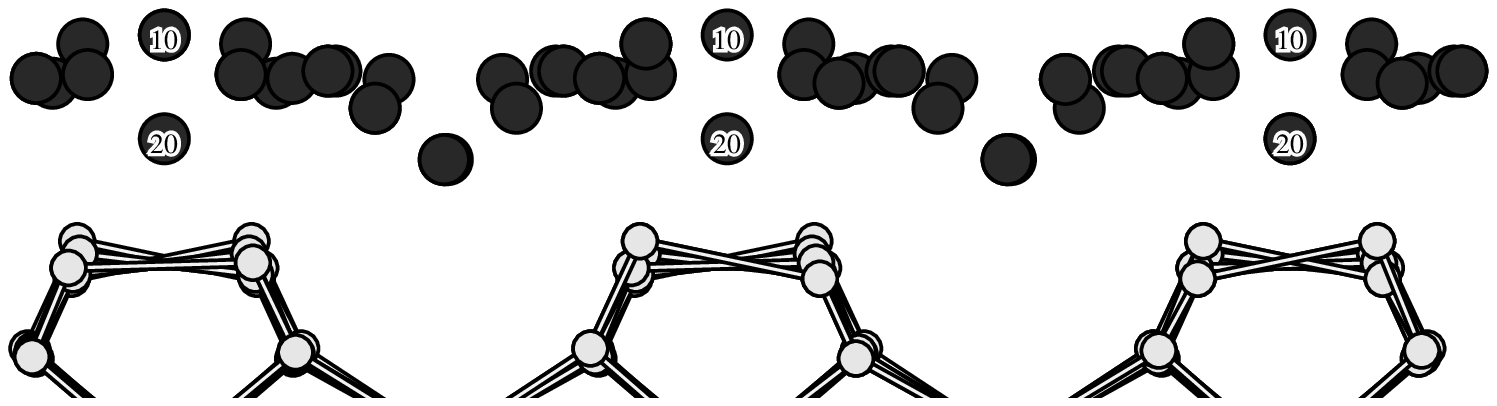}
  \end{center}
  \caption{Atomic structure of the Ge(001)-{\toos}-Pb low-temperature 
    phase. (a) Top-view. The {\toos} unit-cell is indicated by a dashed line. 
    Lead atoms are drawn black or dark grey. 
    For the dark grey lead atoms the distance to the 
    substrate is in the range from 2.6~\AA{} to 2.8~\AA{} and indicates 
    a partially covalent bond. 
    Pb-Pb distances of 3.0~\AA{} are indicated by a dotted line. 
    Germanium atoms are drawn as small circles. 
    The dark grey germanium atoms have a lower $z$-position indicating 
    sp$^2$-like hybridization. 
    The numbers shown correspond to \tref{tab:2106-1} and \tref{tab:2106-2}. 
    Lead atoms 10 and 20 are on high-symmetry sites. 
    (b) Side-view to show the height difference of the lead atoms 10 and 20 
    and the germanium dimer buckling. 
    \label{fig:2106-model}}
\end{figure}

\begin{figure}
  \begin{center}
    \includegraphics[width=9cm]{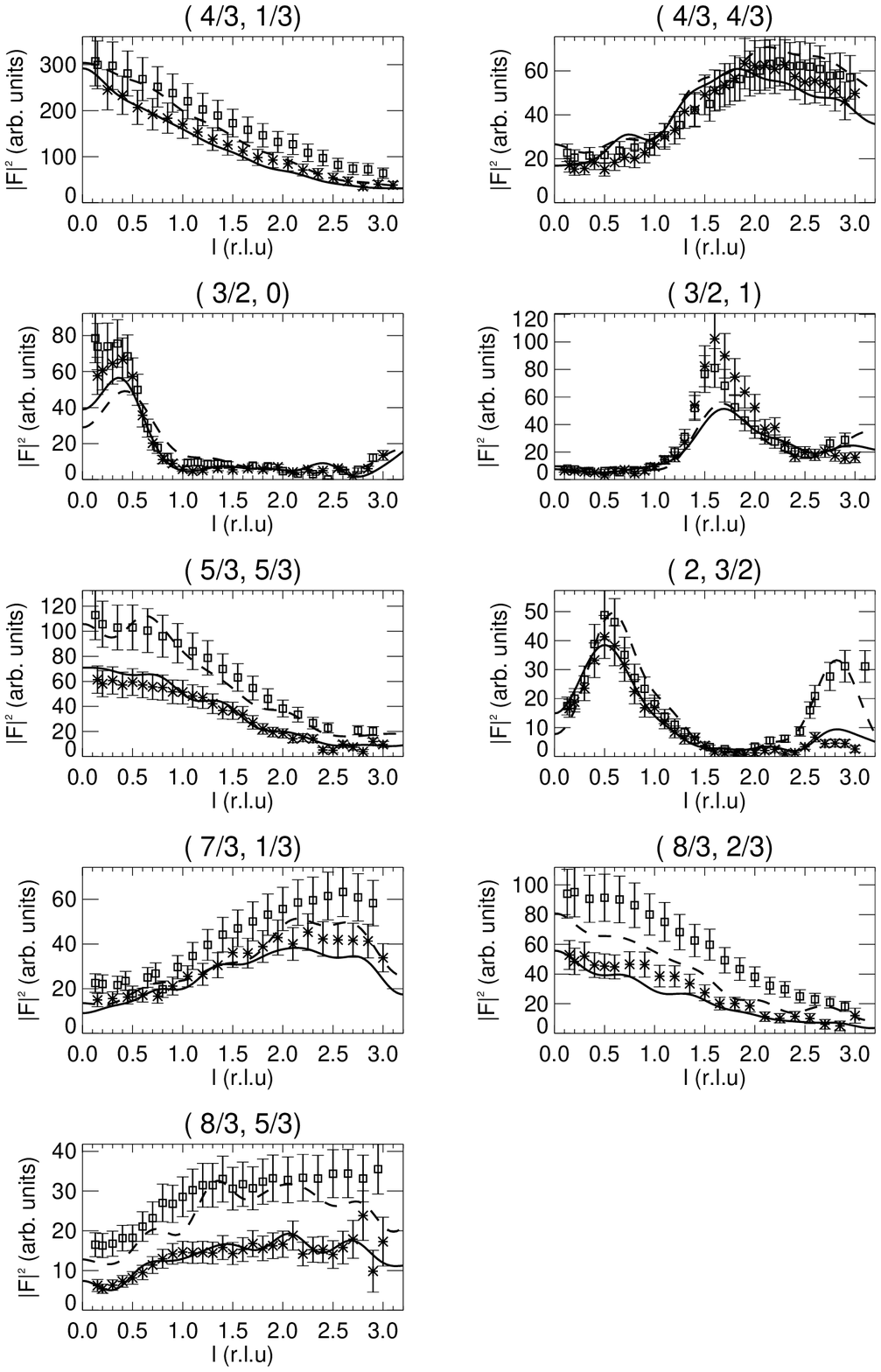}
  \end{center}
  \caption{Comparison of some fractional-order rod scans for the 
    Ge(001)-{\toot}-Pb room-temperature phase 
    (data points: asterisks, calculated intensity: solid line) 
    and the Ge(001)-{\toos}-Pb low-temperature phase 
    (data points: squares, calculated intensity: dashed line). 
    Despite the overall similarities there are some significant differences, 
    e.g., for high $l$-values along the (2,3/2,$l$)-rod and for all 
    measured $l$-values along the (8/3,5/3,$l$)-rod. 
    \label{fig:frac-rods-cmp}}
\end{figure}

Apart from the additional reflections characteristic of the 
{\toos}-periodicity the in-plane data measured at room and at
low temperature look quite similar 
(see \fref{fig:2103-data}(a) and \fref{fig:2106-data}(a)) --
mainly due to the characteristic ring pattern formed by the 
strongest reflections. 
This indicates that the fundamental framework of the
reconstruction remains intact upon cooling. 
But despite the similarities there are definite differences as 
can be seen by the comparison of the fractional-order rod scans shown in 
\fref{fig:frac-rods-cmp}. The differences are not artifacts caused 
by choosing incorrect scaling factors since they are not uniform along 
all rods and they are not just caused by changes in thermal motion 
since they occur in some cases only in a certain range of
$l$-values (see, e.g., the (2,3/2,$l$)-rod for $l > 2$) 
and in other cases along the 
whole rod (see, e.g., the (8/3, 5/3, $l$)-rod). 
The parameters determined in the detailed data analysis are 
listed in \tref{tab:2106-1} and \tref{tab:2106-2} 
and the model is shown in \fref{fig:2106-model}. 
The weakly bonded lead atom (number 10 in \fref{fig:2103-model}) 
remains in a high-symmetry site upon cooling (numbers 10 and 20 in 
\fref{fig:2106-model}). The differences to a similar structural 
model for the low-temperature {\toos}-reconstruction with lead 
atom 2 on a high-symmetry trench site amounts only up to 0.09~\AA{} 
for most atoms. 
But the height (i.e.\ $z$-position) difference of 1.48~\AA{} for 
the high-symmetry site atoms 10 and 20 is necessary to describe 
the data adequately and therefore the chosen symmetry is unambiguously 
the one occurring in the low-temperature phase. 
For the lead layer the lowering of the $z$-position of atom 20, causing 
the described height difference, and the outward displacement of
the next neighbor lead atoms 16 and 19 by 0.77~\AA{} are the most
pronounced changes in comparison 
to the room-temperature structure. All lead atoms are adequately described by 
one in-plane and one out-of-plane ADP. The ADP amplitudes of 
$0.21\pm 0.03$~\AA{} in the surface plane are smaller 
than the $0.26\pm 0.04$~\AA{} and $0.34\pm 0.10$~\AA{} at room temperature, 
but not as low as one would expect solely from a thermal effect. 
Static disorder would explain the larger in-plane
ADP-amplitudes. 
The counterpart of the ``floating'' lead overlayer at room 
temperature, which is only weakly pinned to its commensurate 
position, is probably a network made up of a variety of very similar 
low-temperature structures, which forms due to a broad minimum 
of the total energy as a function of the in-plane position. 

The germanium atoms in the low-temperature phase are well described by 
an isotropic ADP with a value close to zero. This is clear evidence of 
a well-ordered substrate structure at low temperatures. 
The precise value of this 
ADP does not strongly influence the reduced $\chi^2$ and therefore it 
was fixed at a constant small value. 

The germanium dimers close to the lead atoms 10 and 20 are statically 
tilted in alternate directions and probably the 
dynamical motion is frozen out at low temperatures. 
The germanium dimer atoms next to lead 
atom 10 have a $z$-position difference of $0.29\pm 0.06$~\AA{}. 
The germanium dimer atoms next to lead atom 20, which has a significantly 
lower $z$-position than at room temperature, have a larger 
$z$-position difference of $0.54\pm 0.05$~\AA{}. 
Both values are smaller than the corresponding values for the clean surface: 
$0.74 \pm 0.15$~\AA{}\cite{rossmann_ge0012x1_surfsci:1992} 
and $0.69 \pm 0.04$~\AA{} \cite{torrelles_ge0012x1_surfsci:1996} 
for the room-temperature phase 
and $0.84 \pm 0.02$~\AA{} \cite{ferrer_ge0012x1_prl:1995} 
for the low-temperature phase. 
This indicates a stronger interaction of the adsorbate with the substrate 
at low temperature than at room temperature. 
In addition the dimers 21-27 and 
30-32, which probably have a partially covalent interaction 
with the lead layer at room temperature are (within the uncertainty of 
the measurements) symmetric at low-temperature. 
A symmetrical configuration 
would be energetically unfavourable without germanium-lead interactions 
arising from the adsorbate layer. 
It is interesting that not only the distance from these germanium atoms 
in a symmetric configuration to their next neighbor lead atoms is close 
to the covalent bond distance, but also the approximately sp$^3$ hybridized 
germanium atoms 26 and 31 of the statically tilted dimers have a 
distance similar to the covalent bond length to the lead atoms 1 and 3. 
Therefore, we conclude that the lead layer in the low-temperature phase is 
more firmly bonded to the substrate with a more covalent character 
than it is the case at room temperature. 

\begin{figure}
  \begin{center}
    \includegraphics[width=7cm]{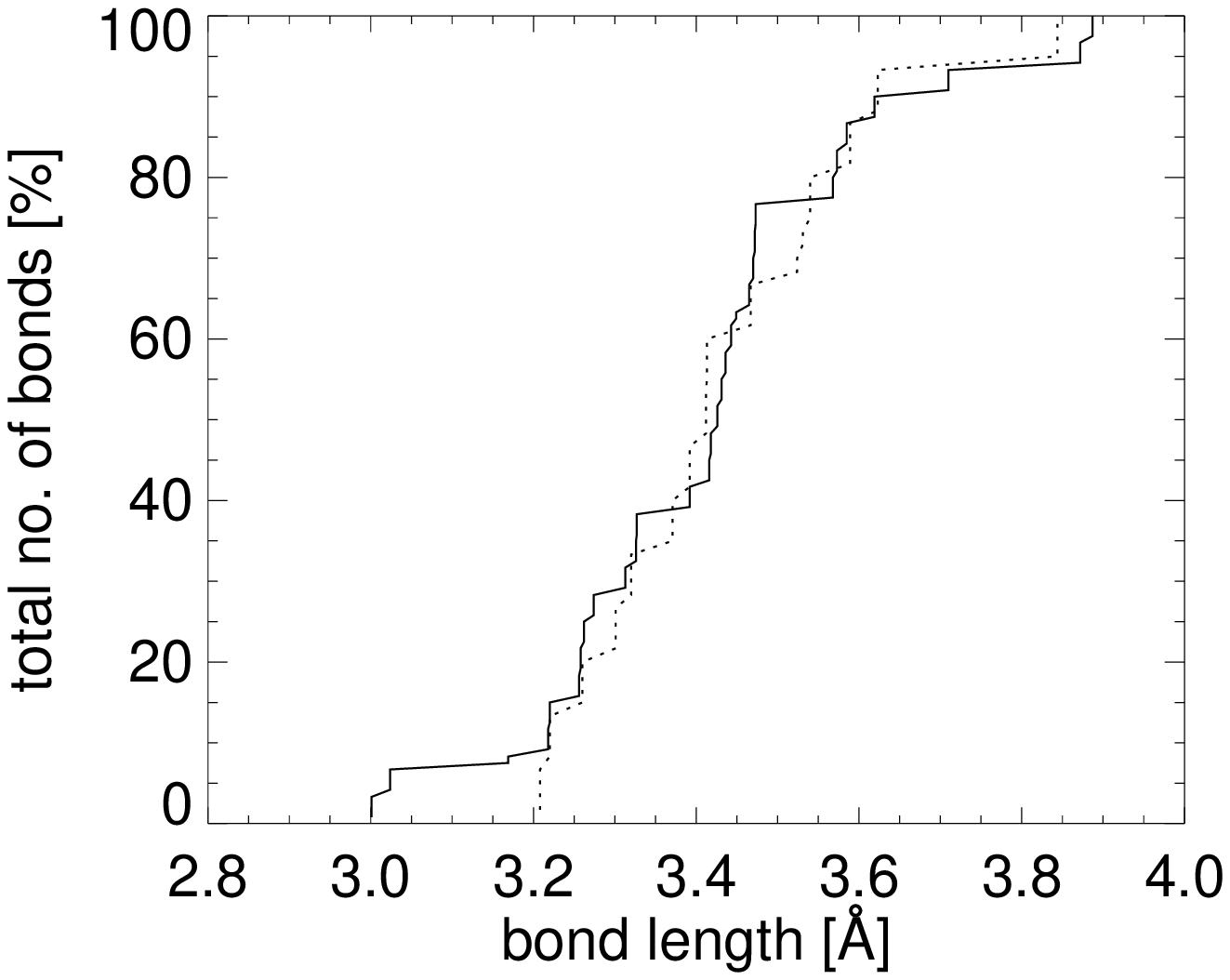}
  \end{center}
  \caption{
    Number of lead atoms with a next neighbor distance 
    within the lead layer below a certain threshold value versus the 
    threshold value distance for the Ge(001)-{\toot}-Pb room-temperature 
    (dashed line) and the Ge(001)-{\toos}-Pb low-temperature phase 
    (solid line). There is a pronounced difference in the region from 
    3.0~{\AA} to 3.2~{\AA} indicating stronger interaction between the 
    lead atoms in the low-temperature phase than in the room-temperature 
    phase. The interaction may have partially covalent character. 
    \label{fig:bond-lengths}}
\end{figure}

The covalent bond length for lead is 2.94~\AA{}. The lead atom 
pairs (1,6), (3,9), (11,16) and (13,19) have a distance of 
$3.0 \pm 0.05$~\AA{} between the atoms of each pair. This may indicate 
the presence of partially covalent bonds even within the overlayer. 
This assumption is supported by \fref{fig:bond-lengths}. 
In this figure the number of lead atoms with next neighbor distances 
(within the lead layer) below a certain threshold value versus the 
threshold value distance is shown. The only significant difference 
between the curves for the room and low-temperature phase 
(dashed and solid line respectively) is a plateau from 
$\sim 3.0$~\AA{} to $\sim 3.2$~\AA{} for the low-temperature phase curve. 
That means the small distances 
within the lead layer are a significant feature 
of the low-temperature phase. 
It is interesting to note that the lead atom pairs with small 
interatomic distances surround the lead atoms 10 and 20, 
which exhibit a pronounced height difference compared to 
their position at room temperature. 

Assuming that there are partially covalent Ge-Pb and Pb-Pb bonds 
immediately raises the question of the hybridization of the lead atoms and 
the electron density distribution within the lead layer. 
As a first approximation one may assume that one germanium atom of each 
buckled dimer has approximately two electrons for a Pb-Ge bond and that the 
electron density within the lead layer is not strongly influenced by these
bonds. 
Each germanium atom of the symmetric dimers has approximately one electron 
and the lead overlayer has to donate one electron for each 
covalent Pb-Ge bond. 
A completely covalent Pb-Pb bond would localize two electrons. 
Therefore, covalent bonds from symmetric germanium dimers to lead atoms 
and within the lead layer will lower the 
electron density in other parts of the overlayer which means that 
the number of partially 
covalent bonds in the energetically most favorable configuration is 
limited. The formation of stronger covalent bonds is inherently 
inhibited by the basic atomic structure of the system; the pathway for 
lowering the total energy of the system by the formation of covalent bonds 
is inherently frustrated. 
The only way out of this frustration would be the formation of a completely 
new structure with a lower total energy. However no such structure 
with a lead coverage of 1.5~ML has been observed and 
it is unlikely that the activation energy for this phase transition 
involving considerable mass transport would be available at low temperatures. 

Considering the classification of the phase transition the results of 
the structure determination presented here suggest that it is of 
order-disorder type. 
The low-temperature phase with static germanium dimers and stronger 
bonding is the ordered form of the room-temperature phase with 
dynamically flipping dimers and a ``floating'' lead overlayer. 
We propose that within the Pb/Ge(001) system there 
is a delicate balance of different pathways for lowering the total energy 
of the system and that the true nature of the system is probably 
more complex than our simplistic bonding arguments would suggest. 

It is instructive to compare the phase transition described here with the 
transition from the hexagonal incommensurate phase of lead on Si(111) 
to the Si(111)-{\ttmoo}-Pb low-temperature structure 
\cite{kumpf_si111pb_surfsci:2000}. 
In both systems the adsorbate forms a Pb(111) overlayer. 
The overlayer is incommensurate (Pb/Si(111)) or weakly pinned to 
a commensurate position (Pb/Ge(001)). 
In both low-temperature phases interatomic distances indicate the 
presence of Pb-substrate bonds with predominantly covalent character. 
The major differences arise from the substrate; in the case of Si(111)
the lowest energy structure, the (7$\times$7) reconstruction, 
is removed by lead deposition and annealing -- 
the silicon substrate of the hexagonal incommensurate phase 
has essentially the undisturbed bulk-truncated structure 
which is not stable without the lead overlayer. 
In the case of Ge(001) the dimers in the (2$\times$1)-reconstruction 
of the clean surface are partially removed by lead deposition 
and reestablished themselves at larger lead coverages.
Therefore, 
the lead overlayer of the Ge(001)-{\toot}-Pb reconstruction stabilizes the 
reconstructed substrate and the lead overlayer of the hexagonal incommensurate 
phase on Si(111) stabilizes the bulk-truncated structure of the substrate. 
Both systems lower their total energy upon cooling by the formation of 
bonds with covalent character and the capability of forming 
such bonds plays a key role in determining the substrate structure. 

\section{Summary}

At room temperature the topmost germanium layer of the 
Ge(001)-{\toot}-Pb recon\-struction is similar to the clean 
(2$\times$1) reconstructed surface which is 
described by dynamically flipping germanium dimers. 
The mainly metallic lead overlayer is probably pinned 
to a commensurate registry with the substrate by weak covalent 
interactions to one germanium dimer per unit-cell. 

Upon cooling, the dimers with a weak covalent bond 
to the adsorbate at room temperature 
remain in a symmetrical position. 
The remaining dimers are statically tilted in alternate directions 
as in the Ge(001)-(2$\times$1) surface reconstruction at low-temperatures. 
Within the lead overlayer the main difference 
to the positions at room temperature is a height difference of 
the two lead atoms that are weakly bonded on high-symmetry sites 
at room temperature. 
Additional partially covalent interactions between the substrate 
and adsorbate and within the adsorbate layer stabilize the system. 

The structural changes are consistent with the occurrence 
of an order-disorder phase transition in this system. 
The phase transition is partially frustrated by the mutual 
exclusiveness of different pathways for lowering the total energy. 

Total energy calculations for this complex system with such a large unit-cell 
are definitely a challenge but, with regard to the interesting properties 
and the expected insight in the delicate energy balance, 
probably worth the effort. We hope that the determination of the detailed 
atomic structure reported here will pave the way for 
theoretical investigations of the Ge(001)-{\toos}-Pb reconstruction 
at $T=0$~K and at elevated temperatures. 

\ack
We thank the staff of HASYLAB for their technical assistance.
Financial support from the Danish Research Council through Dansync,
the Bundesministerium f\"ur Bildung, Wissenschaft,
Forschung und Technologie (BMBF) under project no. 05~KS1GUC/3, 
the ``Graduiertenkolleg Physik nanostrukturierter Festk\"orper'' and the
Volks\-wagen Stiftung is gratefully acknowledged.


~\pagebreak

\Tables

\begin{table}
  \begin{indented}\tiny
  \lineup
  \item[]\begin{tabular}{@{}lllllll}
    \br
no. & type & pos. [surf.~coord.]& \mc{dev.~[{\AA}]} & \mc{ADP $xy/z$ [{\AA}]}\\
    \mr
 1&Pb&(0.740,0.756, 0.444)& \mc{}   &$0.26\pm0.04$&$0.03\pm0.06$\\
 2&Pb&(1.500,3.500, 0.179)& \mc{}   &$0.26\pm0.04$&$0.03\pm0.06$\\
 3&Pb&(0.260,2.244, 0.444)& \mc{}   &$0.26\pm0.04$&$0.03\pm0.06$\\
 4&Pb&(1.067,2.084, 0.401)& \mc{}   &$0.26\pm0.04$&$0.03\pm0.06$\\
 5&Pb&(1.933,1.916, 0.401)& \mc{}   &$0.26\pm0.04$&$0.03\pm0.06$\\
 6&Pb&(1.297,1.300, 0.300)& \mc{}   &$0.26\pm0.04$&$0.03\pm0.06$\\
 7&Pb&(0.066,0.073, 0.405)& \mc{}   &$0.26\pm0.04$&$0.03\pm0.06$\\
 8&Pb&(0.934,2.927, 0.405)& \mc{}   &$0.26\pm0.04$&$0.03\pm0.06$\\
 9&Pb&(1.703,2.700, 0.300)& \mc{}   &$0.26\pm0.04$&$0.03\pm0.06$\\
10&Pb&(0.500,1.500, 0.569)& \mc{}   &$0.34\pm0.10$&$3.0~\pm1.6$\\
\mr
11&Ge&(0.834,3.006,-0.066)&(-0.66, 0.02,-0.37)&0.76&$0.08\pm0.05$&$0.4\pm0.1$\\
12&Ge&(0.201,2.002,-0.064)&( 0.81, 0.01,-0.36)&0.88&$0.08\pm0.05$&$0.4\pm0.1$\\
13&Ge&(0.813,2.015,-0.069)&(-0.75, 0.06,-0.39)&0.85&$0.08\pm0.05$&$0.4\pm0.1$\\
14&Ge&(0.187,0.985,-0.069)&( 0.75,-0.06,-0.39)&0.85&$0.08\pm0.05$&$0.4\pm0.1$\\
15&Ge&(0.799,0.998,-0.064)&(-0.81,-0.01,-0.36)&0.88&$0.08\pm0.05$&$0.4\pm0.1$\\
16&Ge&(0.166,2.994,-0.066)&( 0.66,-0.02,-0.37)&0.76&$0.08\pm0.05$&$0.4\pm0.1$\\
\mr
17&Ge&(1.500,3.500,-0.532)&( 0.00, 0.00,-0.18)&0.18&\mc{$0.13\pm0.06$}\\
18&Ge&(0.031,2.498,-0.291)&( 0.12,-0.01,-0.23)&0.26&\mc{$0.13\pm0.06$}\\
19&Ge&(0.496,2.501,-0.574)&(-0.02, 0.01,-0.42)&0.42&\mc{$0.13\pm0.06$}\\
20&Ge&(0.967,2.507,-0.298)&(-0.13, 0.03,-0.27)&0.30&\mc{$0.13\pm0.06$}\\
21&Ge&(1.506,2.499,-0.521)&( 0.03,-0.00,-0.12)&0.12&\mc{$0.13\pm0.06$}\\
22&Ge&(0.037,1.499,-0.281)&( 0.15,-0.01,-0.17)&0.23&\mc{$0.13\pm0.06$}\\
23&Ge&(0.500,1.500,-0.559)&( 0.00, 0.00,-0.33)&0.33&\mc{$0.13\pm0.06$}\\
24&Ge&(0.963,1.501,-0.281)&(-0.15, 0.01,-0.17)&0.23&\mc{$0.13\pm0.06$}\\
25&Ge&(1.494,1.501,-0.521)&(-0.03, 0.00,-0.12)&0.12&\mc{$0.13\pm0.06$}\\
26&Ge&(0.033,0.493,-0.298)&( 0.13,-0.03,-0.27)&0.30&\mc{$0.13\pm0.06$}\\
27&Ge&(0.504,0.499,-0.574)&( 0.02,-0.01,-0.42)&0.42&\mc{$0.13\pm0.06$}\\
28&Ge&(0.969,0.502,-0.291)&(-0.12, 0.01,-0.23)&0.26&\mc{$0.13\pm0.06$}\\
\mr
29&Ge&(1.993,1.008,-1.024)&(-0.03, 0.03,-0.14)&0.14&\mc{0.08}\\
30&Ge&(0.500,3.000,-0.794)&( 0.00, 0.00,-0.25)&0.25&\mc{0.08}\\
31&Ge&(1.007,2.992,-1.024)&( 0.03,-0.03,-0.14)&0.14&\mc{0.08}\\
32&Ge&(1.514,2.992,-0.774)&( 0.06,-0.03,-0.14)&0.15&\mc{0.08}\\
33&Ge&(1.997,3.001,-1.032)&(-0.01, 0.00,-0.18)&0.18&\mc{0.08}\\
34&Ge&(0.501,1.998,-0.796)&( 0.00,-0.01,-0.26)&0.26&\mc{0.08}\\
35&Ge&(1.018,1.999,-1.037)&( 0.07,-0.01,-0.21)&0.22&\mc{0.08}\\
36&Ge&(1.500,2.000,-0.769)&( 0.00, 0.00,-0.11)&0.11&\mc{0.08}\\
37&Ge&(1.982,2.001,-1.037)&(-0.07, 0.01,-0.21)&0.22&\mc{0.08}\\
38&Ge&(0.499,1.002,-0.796)&(-0.00, 0.01,-0.26)&0.26&\mc{0.08}\\
39&Ge&(1.003,0.999,-1.032)&( 0.01,-0.00,-0.18)&0.18&\mc{0.08}\\
40&Ge&(1.486,1.008,-0.774)&(-0.06, 0.03,-0.14)&0.15&\mc{0.08}\\
\mr
41&Ge&(0.520,0.500,-1.509)&( 0.08,-0.00,-0.05)&0.10&\mc{0.08}\\
42&Ge&(1.012,3.493,-1.264)&( 0.05,-0.03,-0.08)&0.10&\mc{0.08}\\
43&Ge&(1.500,3.500,-1.516)&( 0.00, 0.00,-0.09)&0.09&\mc{0.08}\\
44&Ge&(1.988,3.507,-1.264)&(-0.05, 0.03,-0.08)&0.10&\mc{0.08}\\
45&Ge&(0.480,2.500,-1.509)&(-0.08, 0.00,-0.05)&0.10&\mc{0.08}\\
46&Ge&(0.991,2.503,-1.273)&(-0.04, 0.01,-0.13)&0.14&\mc{0.08}\\
47&Ge&(1.488,2.500,-1.517)&(-0.05,-0.00,-0.10)&0.11&\mc{0.08}\\
48&Ge&(1.990,2.510,-1.280)&(-0.04, 0.04,-0.17)&0.18&\mc{0.08}\\
49&Ge&(0.500,1.500,-1.512)&( 0.00, 0.00,-0.07)&0.07&\mc{0.08}\\
50&Ge&(1.010,1.490,-1.280)&( 0.04,-0.04,-0.17)&0.18&\mc{0.08}\\
51&Ge&(1.512,1.500,-1.517)&( 0.05, 0.00,-0.10)&0.11&\mc{0.08}\\
52&Ge&(0.009,0.497,-1.273)&( 0.04,-0.01,-0.13)&0.14&\mc{0.08}\\
    \br
  \end{tabular}
  \caption{
    Atomic positions in the Ge(001)-{\toot}-Pb room-temperature 
    phase. 
    The atomic positions determined in the SXRD data analysis are given 
    in surface coordinates. The deviation from a bulk-like position 
    and the atomic displacement parameter amplitudes are given in {\AA}. 
    The uncertainty is estimated to be $\pm 0.04$~{\AA} ($\pm 0.02$~{\AA} in 
    $x$, $y$ and $z$) for the Pb atom positions and $\pm 0.09$~{\AA} 
    ($\pm 0.05$~{\AA} in $x$, $y$ and $z$) for the Ge atom positions. 
    Due to the high ADP in the $z$-direction the uncertainty for the 
    $z$-position of the Pb atom number 10 is around $\pm 0.5$~{\AA}. 
  \label{tab:2103}}
  \end{indented}
\end{table}
~\pagebreak

\begin{table}
  \begin{indented}\tiny
  \lineup
  \item[]\begin{tabular}{@{}lllllll}
    \br
no. & type & pos. [surf.~coord.]& \mc{dev.~[{\AA}]} & \mc{ADP $xy/z$ [{\AA}]}\\
   \mr
1&Pb &(0.773,0.741, 0.448)&( 0.13,-0.06, 0.02)&0.14&$0.21\pm 0.03$&$0.05\pm 0.05$\\
 2&Pb &(1.506,3.485, 0.234)&( 0.02,-0.06, 0.31)&0.32&$0.21\pm 0.03$&$0.05\pm 0.05$\\
 3&Pb &(0.227,2.259, 0.448)&(-0.13, 0.06, 0.02)&0.14&$0.21\pm 0.03$&$0.05\pm 0.05$\\
 4&Pb &(1.082,2.103, 0.457)&( 0.06, 0.08, 0.31)&0.33&$0.21\pm 0.03$&$0.05\pm 0.05$\\
 5&Pb &(1.918,1.897, 0.457)&(-0.06,-0.08, 0.31)&0.33&$0.21\pm 0.03$&$0.05\pm 0.05$\\
 6&Pb &(1.249,1.316, 0.363)&(-0.19, 0.07, 0.36)&0.41&$0.21\pm 0.03$&$0.05\pm 0.05$\\
 7&Pb &(0.046,0.103, 0.438)&(-0.08, 0.12, 0.19)&0.24&$0.21\pm 0.03$&$0.05\pm 0.05$\\
 8&Pb &(0.954,2.897, 0.438)&( 0.08,-0.12, 0.19)&0.24&$0.21\pm 0.03$&$0.05\pm 0.05$\\
 9&Pb &(1.751,2.684, 0.363)&( 0.19,-0.07, 0.36)&0.41&$0.21\pm 0.03$&$0.05\pm 0.05$\\
10&Pb &(0.500,1.500, 0.548)&( 0.00, 0.00,-0.12)&0.12&$0.21\pm 0.03$&$0.05\pm 0.05$\\
11&Pb &(0.789,3.726, 0.525)&( 0.20,-0.12, 0.46)&0.51&$0.21\pm 0.03$&$0.05\pm 0.05$\\
12&Pb &(1.494,6.515, 0.234)&(-0.02, 0.06, 0.31)&0.32&$0.21\pm 0.03$&$0.05\pm 0.05$\\
13&Pb &(0.211,5.274, 0.525)&(-0.20, 0.12, 0.46)&0.51&$0.21\pm 0.03$&$0.05\pm 0.05$\\
14&Pb &(1.110,5.069, 0.456)&( 0.17,-0.06, 0.31)&0.36&$0.21\pm 0.03$&$0.05\pm 0.05$\\
15&Pb &(1.890,4.931, 0.456)&(-0.17, 0.06, 0.31)&0.36&$0.21\pm 0.03$&$0.05\pm 0.05$\\
16&Pb &(1.298,4.263, 0.436)&( 0.00,-0.15, 0.77)&0.78&$0.21\pm 0.03$&$0.05\pm 0.05$\\
17&Pb &(0.103,3.053, 0.426)&( 0.15,-0.08, 0.12)&0.21&$0.21\pm 0.03$&$0.05\pm 0.05$\\
18&Pb &(0.897,5.947, 0.426)&(-0.15, 0.08, 0.12)&0.21&$0.21\pm 0.03$&$0.05\pm 0.05$\\
19&Pb &(1.702,5.737, 0.436)&(-0.00, 0.15, 0.77)&0.78&$0.21\pm 0.03$&$0.05\pm 0.05$\\
20&Pb &(0.500,4.500, 0.287)&( 0.00, 0.00,-1.60)&1.60&$0.21\pm 0.03$&$0.05\pm 0.05$\\
\mr
21&Ge &(0.815,5.995,-0.025)&(-0.08,-0.04, 0.23)&0.25&\mc{0.04}\\
22&Ge &(0.191,5.002, 0.030)&(-0.04, 0.00, 0.53)&0.53&\mc{0.04}\\
23&Ge (sp$^2$)&(0.829,5.012,-0.065)&( 0.07,-0.01, 0.02)&0.07&\mc{0.04}\\
24&Ge (sp$^2$)&(0.171,3.988,-0.065)&(-0.07, 0.01, 0.02)&0.07&\mc{0.04}\\
25&Ge &(0.809,3.998, 0.030)&( 0.04,-0.00, 0.53)&0.53&\mc{0.04}\\
26&Ge &(0.801,1.000,-0.002)&( 0.01, 0.01, 0.35)&0.35&\mc{0.04}\\
27&Ge &(0.160,5.986,-0.038)&(-0.02,-0.03, 0.16)&0.16&\mc{0.04}\\
28&Ge (sp$^2$)&(0.824,2.004,-0.054)&( 0.05,-0.04, 0.08)&0.10&\mc{0.04}\\
29&Ge (sp$^2$)&(0.176,0.996,-0.054)&(-0.05, 0.04, 0.08)&0.10&\mc{0.04}\\
30&Ge &(0.840,3.014,-0.038)&( 0.02, 0.03, 0.16)&0.16&\mc{0.04}\\
31&Ge &(0.199,2.000,-0.002)&(-0.01,-0.01, 0.35)&0.35&\mc{0.04}\\
32&Ge &(0.185,3.005,-0.025)&( 0.08, 0.04, 0.23)&0.25&\mc{0.04}\\
\mr
33&Ge &(1.494,6.500,-0.482)&(-0.02, 0.00, 0.28)&0.28&\mc{0.04}\\
34&Ge &(0.014,5.482,-0.240)&(-0.07,-0.06, 0.29)&0.30&\mc{0.04}\\
35&Ge &(0.478,5.489,-0.526)&(-0.07,-0.05, 0.27)&0.29&\mc{0.04}\\
36&Ge &(0.958,5.524,-0.265)&(-0.04, 0.07, 0.18)&0.20&\mc{0.04}\\
37&Ge &(1.494,5.512,-0.475)&(-0.05, 0.05, 0.26)&0.27&\mc{0.04}\\
38&Ge &(0.033,4.503,-0.245)&(-0.02, 0.02, 0.20)&0.20&\mc{0.04}\\
39&Ge &(0.500,4.500,-0.535)&( 0.00, 0.00, 0.14)&0.14&\mc{0.04}\\
40&Ge &(0.967,4.497,-0.245)&( 0.02,-0.02, 0.20)&0.20&\mc{0.04}\\
41&Ge &(1.506,4.488,-0.475)&( 0.05,-0.05, 0.26)&0.27&\mc{0.04}\\
42&Ge &(0.042,3.476,-0.265)&( 0.04,-0.07, 0.18)&0.20&\mc{0.04}\\
43&Ge &(1.513,1.487,-0.479)&( 0.08,-0.06, 0.24)&0.26&\mc{0.04}\\
44&Ge &(0.031,0.471,-0.261)&(-0.01,-0.09, 0.21)&0.22&\mc{0.04}\\
45&Ge &(0.482,0.481,-0.535)&(-0.09,-0.07, 0.22)&0.25&\mc{0.04}\\
46&Ge &(0.967,0.502,-0.250)&(-0.01, 0.00, 0.23)&0.23&\mc{0.04}\\
47&Ge &(1.487,2.513,-0.479)&(-0.08, 0.06, 0.24)&0.26&\mc{0.04}\\
48&Ge &(0.030,1.503,-0.253)&(-0.03, 0.02, 0.16)&0.16&\mc{0.04}\\
49&Ge &(0.499,1.499,-0.531)&(-0.00,-0.01, 0.16)&0.16&\mc{0.04}\\
50&Ge &(0.970,1.497,-0.253)&( 0.03,-0.02, 0.16)&0.16&\mc{0.04}\\
51&Ge &(1.506,3.500,-0.482)&( 0.02,-0.00, 0.28)&0.28&\mc{0.04}\\
52&Ge &(0.033,2.498,-0.250)&( 0.01,-0.00, 0.23)&0.23&\mc{0.04}\\
53&Ge &(0.518,2.519,-0.535)&( 0.09, 0.07, 0.22)&0.25&\mc{0.04}\\
54&Ge &(0.969,2.529,-0.261)&( 0.01, 0.09, 0.21)&0.22&\mc{0.04}\\
55&Ge &(0.522,3.511,-0.526)&( 0.07, 0.05, 0.27)&0.29&\mc{0.04}\\
56&Ge &(0.986,3.518,-0.240)&( 0.07, 0.06, 0.29)&0.30&\mc{0.04}\\
\br
  \end{tabular}
  \caption{Atomic positions in the Ge(001)-{\toos}-Pb low-temperature 
    phase (lead atoms and substrate layers 1-3). 
    The atomic positions determined in the SXRD data analysis are given 
    in surface coordinates. The deviation from a bulk-like position 
    and the atomic displacement parameter amplitudes are given in {\AA}. 
    The uncertainty is estimated to be $\pm 0.06$~{\AA} ($\pm 0.03$~{\AA} in 
    $x$, $y$ and $z$) for the Pb atom positions and $\pm 0.1$~{\AA} 
    (up to $\pm 0.06$~{\AA} in $x$, $y$ and $z$) for the Ge atom positions. 
  \label{tab:2106-1}}
  \end{indented}
\end{table}
~\pagebreak

\begin{table}
  \begin{indented}\tiny
  \lineup
  \item[]\begin{tabular}{@{}lllllll}
    \br
no. & type & pos. [surf.~coord.]& \mc{dev.~[{\AA}]} & \mc{ADP $xy/z$ [{\AA}]}\\
    \mr
 57&Ge &(1.997,1.005,-1.001)&( 0.02,-0.01, 0.13)&0.14&\mc{0.04}\\
 58&Ge &(0.500,5.990,-0.775)&(-0.00,-0.04, 0.11)&0.12&\mc{0.04}\\
 59&Ge &(1.004,5.986,-1.004)&(-0.01,-0.02, 0.11)&0.12&\mc{0.04}\\
 60&Ge &(1.494,5.998,-0.741)&(-0.08, 0.02, 0.19)&0.21&\mc{0.04}\\
 61&Ge &(1.989,6.017,-1.001)&(-0.03, 0.07, 0.18)&0.19&\mc{0.04}\\
 62&Ge &(0.497,5.000,-0.778)&(-0.02, 0.01, 0.10)&0.10&\mc{0.04}\\
 63&Ge &(1.021,5.002,-0.999)&( 0.01, 0.02, 0.21)&0.21&\mc{0.04}\\
 64&Ge &(1.500,5.000,-0.726)&( 0.00, 0.00, 0.25)&0.25&\mc{0.04}\\
 65&Ge &(1.979,4.998,-0.999)&(-0.01,-0.02, 0.21)&0.21&\mc{0.04}\\
 66&Ge &(0.503,4.000,-0.778)&( 0.02,-0.01, 0.10)&0.10&\mc{0.04}\\
 67&Ge &(1.987,2.008,-1.005)&( 0.02, 0.02, 0.18)&0.19&\mc{0.04}\\
 68&Ge &(0.494,1.002,-0.773)&(-0.02,-0.00, 0.13)&0.14&\mc{0.04}\\
 69&Ge &(1.008,0.985,-1.006)&( 0.02,-0.06, 0.15)&0.16&\mc{0.04}\\
 70&Ge &(1.498,0.992,-0.745)&( 0.05,-0.06, 0.16)&0.18&\mc{0.04}\\
 71&Ge &(1.992,3.015,-1.006)&(-0.02, 0.06, 0.15)&0.16&\mc{0.04}\\
 72&Ge &(0.506,1.998,-0.773)&( 0.02, 0.00, 0.13)&0.14&\mc{0.04}\\
 73&Ge &(1.013,1.992,-1.005)&(-0.02,-0.02, 0.18)&0.19&\mc{0.04}\\
 74&Ge &(1.500,2.000,-0.734)&( 0.00, 0.00, 0.20)&0.20&\mc{0.04}\\
 75&Ge &(1.996,4.014,-1.004)&( 0.01, 0.02, 0.11)&0.12&\mc{0.04}\\
 76&Ge &(0.500,3.010,-0.775)&( 0.00, 0.04, 0.11)&0.12&\mc{0.04}\\
 77&Ge &(1.003,2.995,-1.001)&(-0.02, 0.01, 0.13)&0.14&\mc{0.04}\\
 78&Ge &(1.502,3.008,-0.745)&(-0.05, 0.06, 0.16)&0.18&\mc{0.04}\\
 79&Ge &(1.011,3.983,-1.001)&( 0.03,-0.07, 0.18)&0.19&\mc{0.04}\\
 80&Ge &(1.506,4.002,-0.741)&( 0.08,-0.02, 0.19)&0.21&\mc{0.04}\\
\mr
 81&Ge &(0.477,0.493,-1.499)&(-0.17,-0.03, 0.06)&0.18&\mc{0.04}\\
 82&Ge &(1.003,6.489,-1.257)&(-0.04,-0.02, 0.04)&0.06&\mc{0.04}\\
 83&Ge &(1.504,6.501,-1.501)&( 0.02, 0.00, 0.09)&0.09&\mc{0.04}\\
 84&Ge &(1.996,6.506,-1.252)&( 0.03,-0.00, 0.06)&0.07&\mc{0.04}\\
 85&Ge &(0.505,5.502,-1.498)&( 0.10, 0.01, 0.06)&0.12&\mc{0.04}\\
 86&Ge &(1.016,5.498,-1.255)&( 0.10,-0.02, 0.10)&0.15&\mc{0.04}\\
 87&Ge &(1.514,5.501,-1.499)&( 0.10, 0.01, 0.10)&0.14&\mc{0.04}\\
 88&Ge &(1.993,5.510,-1.241)&( 0.01,-0.00, 0.22)&0.22&\mc{0.04}\\
 89&Ge &(0.500,4.500,-1.501)&( 0.00, 0.00, 0.06)&0.06&\mc{0.04}\\
 90&Ge &(1.007,4.490,-1.241)&(-0.01, 0.00, 0.22)&0.22&\mc{0.04}\\
 91&Ge &(0.505,1.503,-1.501)&( 0.02, 0.01, 0.06)&0.07&\mc{0.04}\\
 92&Ge &(1.007,1.490,-1.250)&(-0.01, 0.00, 0.17)&0.17&\mc{0.04}\\
 93&Ge &(1.490,1.496,-1.500)&(-0.09,-0.02, 0.10)&0.13&\mc{0.04}\\
 94&Ge &(1.974,1.495,-1.248)&(-0.14,-0.01, 0.15)&0.20&\mc{0.04}\\
 95&Ge &(0.523,2.507,-1.499)&( 0.17, 0.03, 0.06)&0.18&\mc{0.04}\\
 96&Ge &(1.026,2.505,-1.248)&( 0.14, 0.01, 0.15)&0.20&\mc{0.04}\\
 97&Ge &(1.510,2.504,-1.500)&( 0.09, 0.02, 0.10)&0.13&\mc{0.04}\\
 98&Ge &(1.993,2.510,-1.250)&( 0.01,-0.00, 0.17)&0.17&\mc{0.04}\\
 99&Ge &(0.495,3.498,-1.498)&(-0.10,-0.01, 0.06)&0.12&\mc{0.04}\\
100&Ge &(1.004,3.494,-1.252)&(-0.03, 0.00, 0.06)&0.07&\mc{0.04}\\
101&Ge &(1.496,3.499,-1.501)&(-0.02,-0.00, 0.09)&0.09&\mc{0.04}\\
102&Ge &(1.997,3.511,-1.257)&( 0.04, 0.02, 0.04)&0.06&\mc{0.04}\\
103&Ge &(1.486,4.499,-1.499)&(-0.10,-0.01, 0.10)&0.14&\mc{0.04}\\
104&Ge &(1.984,4.502,-1.255)&(-0.10, 0.02, 0.10)&0.15&\mc{0.04}\\
    \br
  \end{tabular}
  \caption{Atomic positions in the Ge(001)-{\toos}-Pb low-temperature 
    phase (substrate layers 4-7). See \tref{tab:2106-1} for further 
    explanations.
  \label{tab:2106-2}}
  \end{indented}
\end{table}
~\pagebreak

\Figures

\end{document}